# GENERAL EXPRESSION FOR THE CHEMICAL SYSTEM RESPONSE TO EXTERNAL IMPACT

*B. Zilbergleyt*[I]

INTRODUCTION

Objective of the open systems theory is to formalize system interaction with its environment. Chemical transformation within the system tends to achieve either the state of "true" thermodynamic equilibrium (TdE), which is unique in isolated systems with only one transformation [1], or the state of external equilibrium with its environment under impact of external thermodynamic force (TdF) in closed/open systems. In discrete thermodynamics of chemical equilibria (DTd) [2] the key issue is the chemical system shift from TdE, caused by the external force. The shift $\delta\xi$ is related to the extent $\Delta\xi$ of chemical transformation ($\delta$ and $\Delta$ in writing) within the system as $\delta=1-\Delta$. Following the logic of Zeroth principle of thermodynamics, starting point for the chemical system analysis in DTd is always the state of TdE with $\Delta=1$ (and $\delta=0$) by definition. According to Le Chatelier's principle, open system changes its current state to reduce the mismatch with acting against it TdF; the system trends either towards the initial point ($\delta\rightarrow1$, $\delta>0$), or towards the logistic end of chemical transformation, where the process feedstock is exhausted ($\delta<0$).

In previous research we have formalized the system response by presenting the external TdF as power series of $\delta$ (the Le Chatelier's response, LCR) [2]

(1) $\qquad\qquad\qquad\qquad\qquad\qquad \Sigma_{0\rightarrow\pi}w_p\delta_j{}^p = -(1/\alpha_j)F_{je},$

the upper power limit $\pi$ in the series is considered a loosely defined system complexity factor with regards to its response to TdF. Expression in the left hand side (1) is dimensionless; the TdF dimension is energy, dimension of $\alpha_j$ must also be energy. Because values of the weights $w_i$ are unknown, we put all them but $w_0$ initially to unities. Switching the weight $w_0$ between two fixed numbers, 0 or 1, we have obtained two respective formal relationships between $\delta_j$ and reduced by RT internal thermodynamic force (reduced bound affinity [2]) as logistic maps of the chemical system states, one for the strong type at $w_0=0$

(2) $\qquad\qquad\qquad\qquad\qquad \ln[\Pi_j(\eta_j,0)/\Pi_j(\eta_j,\delta_j)] - \tau_j(\delta_j - \delta_j{}^{\pi+1}) = 0,$

and another for the weak type of the system response at $w_0=1$

(3) $\qquad\qquad\qquad\qquad\qquad \ln[\Pi_j(\eta_j,0)/\Pi_j(\eta_j,\delta_j)] - \tau_j(1 - \delta_j{}^{\pi+1}) = 0,$

where $\Pi_j(\eta_j,\delta_j)$ are regular mole fraction products, $\Pi_j(\eta_j,0)$ equals to constant of equilibrium $K_j$. All states, predicted by the maps, correspond to equilibria between internal and external thermodynamic forces, acting against the system. Factor $\tau_j$ is the growth factor, like in bio-population theories (aka "demand for pray" [3]), defining the growth of the system deviation from TdE, its denominator is just RT and its numerator is reverse to the proportionality coefficient from (1), interpreted as $\alpha_j=RT_a$ with a fictitious

---

[I] System Dynamics Research Foundation, Chicago, USA, sdrf@ameritech.net

"alternative temperature" $T_a$ [2]. The difference between the maps is slim. However, it leads to essentially different responses to external TdF, visible in bifurcation diagrams, the graphic solutions to the maps(2) and (3), shown in Fig.1. Infinite set of such diagrams covers either the whole I quadrant of the reference frame in case of δ>0 or the III quadrant if δ<0, building up the chemical system domain of states. Although non-rigorously derived (due to suggestion that all $w_{p\neq0}=1$), previous results were logically and mathematically correct, but the "black and white" presentation of the chemical system responses to external TdF as either the strong or the weak with nothing in between is confusing. Although we cannot eliminate still unknown $w_p$, in this work we are presenting new approximation to the theory with a new auxiliary function and more correct derivation, that leads to more general map of states of chemical systems.

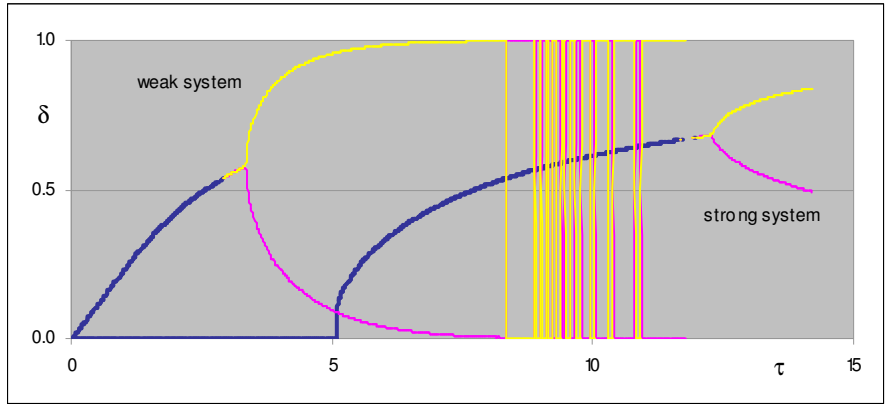

Fig.1. Weak and strong system diagrams, maps (3) and (2), reaction A+B=AB, η=0.689, π=1.

We still use abstract stoichiometric reaction equations and arbitrary standard changes of Gibbs' free energy for more freedom. We also keep using the thermodynamic equivalent of transformation η as major characteristic of the system/reaction robustness, defined in [2] as mole amount of any reaction participant, transformed along the reaction way from the initial point to TdE, per its stoichiometric unit. Having the same value for all participants of a given reaction, it is unambiguously related to the system initial composition and reaction $\Delta G^0$; such relationships for fixed initial composition in direct reaction A+B=AB and its reverse are shown graphically in Fig.2. We use direct reaction to simulate the system response to TdF, driving it towards the reaction reagents, and reverse reaction in case of TdF, driving the system towards the products. Obviously, both half-reactions share the same TdE point, and it was found, that $\eta_d+\eta_r=1$ with the accuracy up to the third decimal sign, depending upon precision in finding the equilibria. One can use this diagram to evaluate $\Delta G^0$ by the value of $\eta$ and vise versa.

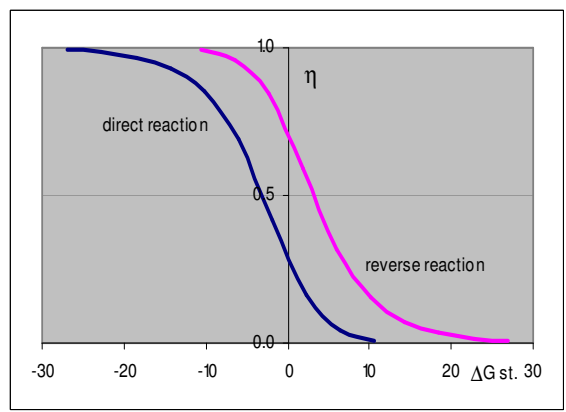

Fig.2. Thermodynamic equivalent of transformation η vs. ΔG⁰, direct reaction
A+B=AB, reagents were taken in stoichiometric relations.

GENERAL MAP FOR THE CHEMICAL SYSTEM RESPONSE

Let's take a look at (1) again. In isolated equilibrium with $F_{je}=0$ we have $\delta_j=0$, $w_0=0$, and (1) may be re-written as

$$(4) \qquad\qquad\qquad \Sigma_{1\to\pi}w_p\delta_j{}^p = -(1/\alpha_j)F_{je}.$$

Again, the weights $w_p$ are unknown *a priori*; it's useful to suppose $w_p \in [0,1]$, but that doesn't eliminate the problem. The situation can be eased, if we introduce another power series approximation of TdF with new weights $\omega_p$

$$(5) \qquad\qquad\qquad \omega_0+\Sigma_{1\to\pi}\delta_j{}^p = -(1/\alpha_j)F_{je},$$

where $\omega_0 \neq 0$, all $\omega_{p\neq 0}=1$, and deduction of (4) from (5) gives

$$(6) \qquad\qquad\qquad \omega_0(\delta_j) - \Sigma_{1\to\pi}(1-w_p)\delta_j{}^p = 0.$$

Now we have an auxiliary function $\omega_0$ to compensate uncertainty in the $w_p$ values. Actual value of $\omega_0$ is defined by the structure of the chemical system response to external impact and obviously falls into range from minimum $\omega_0$ with all $w_p \approx 1$ to maximum $\omega_0$ with all $w_p \approx 0$ and $\delta_j \to 1$, i.e. $0 \leq \omega_0 < \pi$. Now, at $F_{je}=0$ we have $\omega_0=0$ due to $\delta_j=0$, but even a small deviation of the force from zero may lead to visible change of $\omega_0$.

Recalling, that DTd defines equilibrium as a balance between the bound affinity and external thermodynamic forces $A_j+F_{je}=0$, and that in DTd thermodynamic affinity is defined as

$$(7) \qquad\qquad\qquad A_j = -\Delta\Phi_j(\eta_j,\delta_j)_{x,y}/(1-\delta_j),$$

were $\Phi_j$ is appropriate characteristic function, we get

$$(8) \qquad\qquad\qquad \Delta\Phi_j(\eta_j,\delta_j)_{x,y} + (1-\delta_j)F_{je} = 0.$$

At p,T=const with Gibbs' free energy as characteristic function it turns to

$$(9) \qquad\qquad\qquad \Delta G_j(\eta_j,\delta_j) + (1-\delta_j)F_{je} = 0.$$

To obtain the map of states, we substitute $F_{je}$ in (9) by $-\alpha_j\Sigma_{0\to\pi}\omega_p\delta_j{}^p$; unwrap $\Delta G$ as logarithm of molar part products ratio, multiplied by RT; then, after dividing all the expression through by RT and introducing the growth factor $\tau_j=\alpha_j/RT$, we arrive at

$$(10) \qquad\qquad \ln[\Pi_j(\eta_j,0)/\Pi_j(\eta_j,\delta_j)] - \tau_j(1-\delta_j)(\omega_0 + \Sigma_{1\to\pi}\delta_j{}^p) = 0.$$

For restricted $\pi$ (~20 or so) we finally get

$$(11) \qquad\qquad \ln[\Pi_j(\eta_j,0)/\Pi_j(\eta_j,\delta_j)] - \tau_j[\omega_0(1-\delta_j)+\delta_j(1-\delta_j{}^\pi)] = 0.$$

It is easy to see that map (11) turns to map (2) at $\omega_0=0$ and to map (3) at $\omega_0=1$. It is more general map of states of the chemical system, obtained at the price of introduction of new auxiliary function $\omega_0(\delta_j)$; now the uncertainty is located only in that function, and we can easier play around it.

SIMULATION  METHOD  AND SOFTWARE

Map (11) is transcendental, and one has to use numerical methods to find solutions. Simulation software, created by SDRF to solve the DTd problems, runs iterations with $\tau$ as the base. It moves the electronic image of chemical system along the loci of solutions to map (11) and prepares data to get out the processor in $\delta$ vs. $\tau$ coordinates; the results may be recalculated to another reference frame by the user's command. The input includes following information on the system: the chemical reaction parameters - stoichiometric equation, standard change of Gibbs' free energy, thermodynamic temperature; the system parameters - initial amounts of participants and complexity factor $\pi$; iteration parameters - number of "external" iteration steps, t (usually 10,000), defining the iteration step as 1/t, number of "internal" search/iteration steps, $\alpha$ (usually 20-50), and precision $\varepsilon$ in finding zeroes of map (11). First, the software calculates $\eta$, corresponding to $\Delta G^0$, T and initial composition, and populates the array of logarithmic terms of map (11) within the range $(0 < t_i < t+1)$. Iterations start after that by setting next value of $t_i$ and mapping it into the running value of $\tau_i$ ("external" iteration), and then with the step $1/(t \cdot \alpha)$ proceeds to find the $\delta_i$ value, which is the next solution to map (11), or the next point on the locus of the map solutions. The iterations continue until the number of steps exceeds the $\alpha \cdot t$ product. Due to found in our previous works fractality of the chemical system bifurcation diagrams, the oscillations spectra in particular (see below), some details of the diagram shape and its location regarding $\tau$ and $\delta$ axes depend on the size of iteration steps. The software we describe may run through ca 50 million steps in one iteration cycle within a matter of minutes or less, which is more than enough for all feasible tasks.

As it is easy to see, the product of the logarithmic term of map (11) by RT is the system $\Delta G$ in open equilibrium; being divided by $(1-\delta)$, it turns into the system bound thermodynamic affinity in the same state, which is equal to external thermodynamic force, causing the system shift $\delta$ from TdE. That's how the software recalculates the output data.

Each set of input data defines one bifurcation diagram of their infinite amount, constituting the chemical system domain of states.

STATIC SIMULATION  RESULTS

We call static the diagrams in $\tau$-$\delta$ coordinates: no TdF were used to plot them; diagrams in the $F_{je}$-$\delta$ coordinates are dynamic. As it was categorized earlier [2], static bifurcation diagrams of the chemical system response consist of three areas: the TdE area with $\delta_j=0$, where the curve is resting on or very close to abscissa; the area of open equilibrium (OpEq) between the conditional end of TdE area and bifurcation point, both are the parts of thermodynamic branch; and then bi-stable bifurcation area, enveloped by two sub-branches; in some cases this area starts immediately with oscillations. Unlike traditional logistic maps, whose diagrams show instabilities via series of bifurcations with dubbing periods as the growth factor increases further, solutions to map (11) have only one, the period-2 bifurcation point. It is followed by bifurcation area, where stresses, accumulated by the system as external drive increases further, also lead to instabilities, but in form of chaotic oscillations with intricate spectra. We have found them in the systems with reactions, featuring large negative values of $\Delta G^0$ (say, <-15 kJ/m). Static graphic solutions to map (11) for different stoichiometries at maximum $\omega_0$ are shown in Fig.3; all reagents were taken in stoichiometric proportions. Next picture in Fig.4 presents responses of direct reaction to external impact at varied $\omega_0$ with restricted by factor x weights $w_p$. Based on the new model, these results had shown that the TdE area, on one side, and oscillations in the bi-stability zone, on the other side, have been well expressed as extreme "strong" and "weak" cases at the ends of [0,1] interval as well as within it. The system response stays "weak" far beyond $\omega_0=1$, up to $\omega_0 \approx \pi$. New general map (11) unifies most of the encountered types of the system reactions into one class.

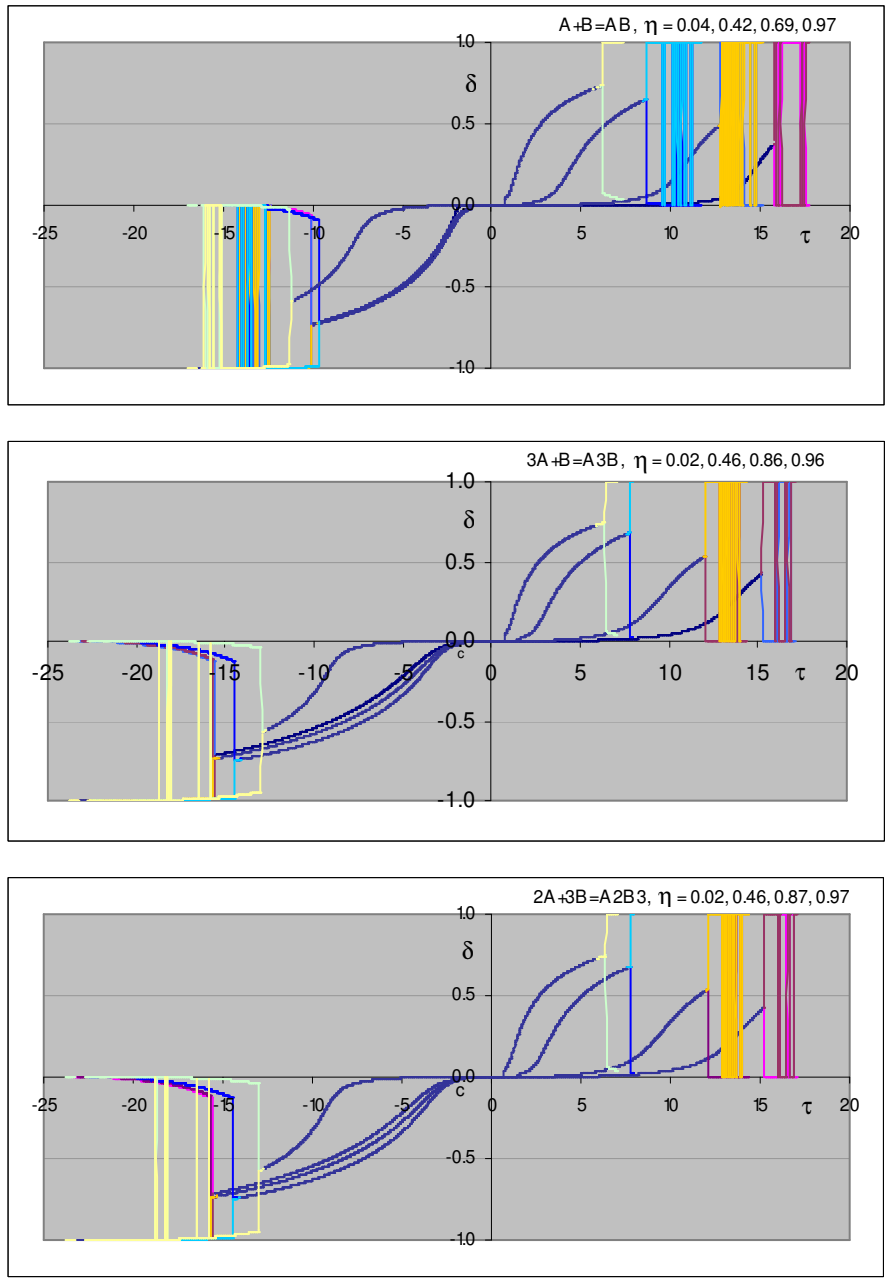

Fig.3. Graphical solutions to map (11), various stoichiometries at maximum potential values of $\omega_0$; the $\eta_d$ values follow the order of the curves; $\pi=1$. The diagrams are similar, but the oscillation spectra are different. The oscillations occur at higher $\eta$.

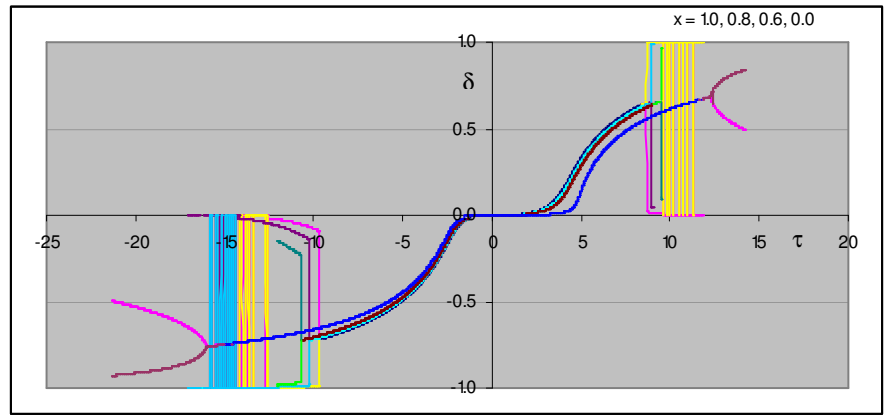

Fig.4. Bifurcation diagrams for the system with direct reaction A+B=AB, $\eta_d$=0.69, $\pi$=1, $\omega_0$=x$\Sigma_{1 \to \pi}$(1−$w_p$)$\delta_j{}^p$, x values follow the curves order, left to the right in the I quadrant.

As one can see in Fig.4, bifurcation diagrams are evolving from one extreme to another as $\omega_0$ sweeps interval [0,1], some curves are degenerated. Solutions to the new map also show an utmost "triggering" shape of bifurcations: sub-branches leave bifurcation point vertically, and bifurcations remind square bracket; to compare with the previous results see Fig.1.

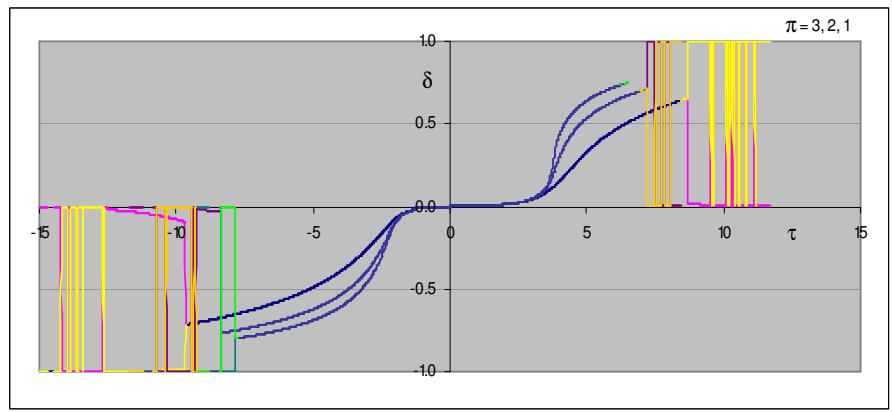

Fig.5. System with reaction A+B=AB, $\eta_d$=0.97, the $\pi$ values follow the curves order, left to the right in the I quadrant.

We also performed simulations to check up the complexity factor influence; the results are in Fig.5. Direct and reverse response graphs are not quite symmetrical regarding the coordinate axes, because turning one into another changes the signs of stoichiometric coefficients and $\Delta G^0$.

DYNAMICS OF THE CHEMICAL SYSTEM RESPONSE

In previous publications we have introduced and made use of dynamic bifurcation diagrams, that show dependence of $\delta$ upon external TdF [2]. It is impossible to produce correctly the output into this reference frame within bifurcation area, and we are forced to restrict their usage by the mono-stable zones. As we mentioned earlier, shift from TdE means changes in the system $\Delta G$, equal to the logarithmic term of map (11), multiplied by RT; being divided by $(1-\delta)$ it gives us the TdF value in kJ/m. Appropriate graphs are shown in Fig.6; all curves on both pictures are ending by bifurcation points.

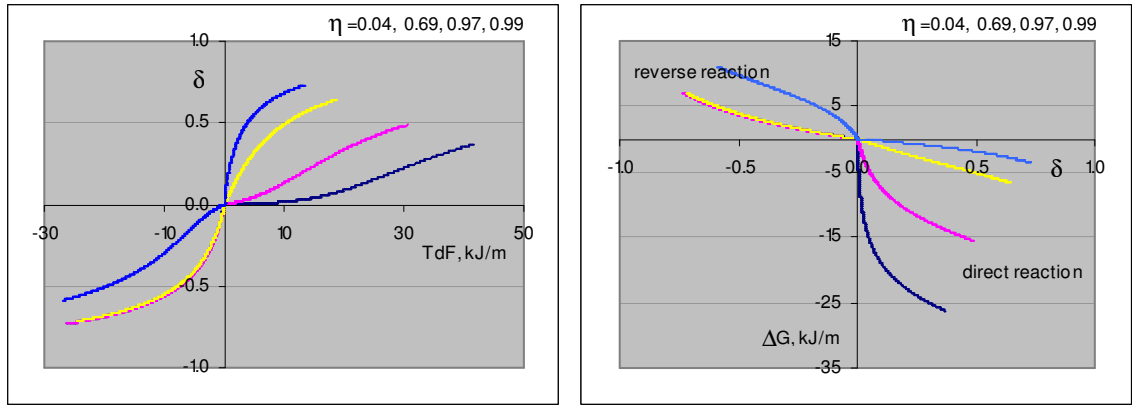

Fig.6. System shift vs. TdF (left), $\eta$ from left to right, and system $\Delta G$ vs. shift (right), $\eta$ from up to down, system with reaction A+B=AB, $\omega_0$ max, $\pi$=1.

FRACTALITY OF THE SOLUTIONS TO THE BASIC MAP

Previous research has revealed a fractal nature of the chaotic oscillations in closed chemical systems [4,5]. With developed in this work new approach to DTd, simulated solutions to map (11) have revealed new features of relevant bifurcation diagrams. Some results, obtained in a trial shot, are shown in Fig.7. Like in Mandelbrot's problem of the Great Britain's coastal line length [6], varying the iteration parameter α in our task has lead to changes in the shape and parameters of thermodynamic branch: the lesser is the internal iteration step (=1/α), the shorter are the TdE and OpE areas, and the shift value at bifurcation point is also going down. The oscillation spectra depend on α as well. The whole domain of states looks as shrinking in a fractal manner when the iteration step gets smaller.

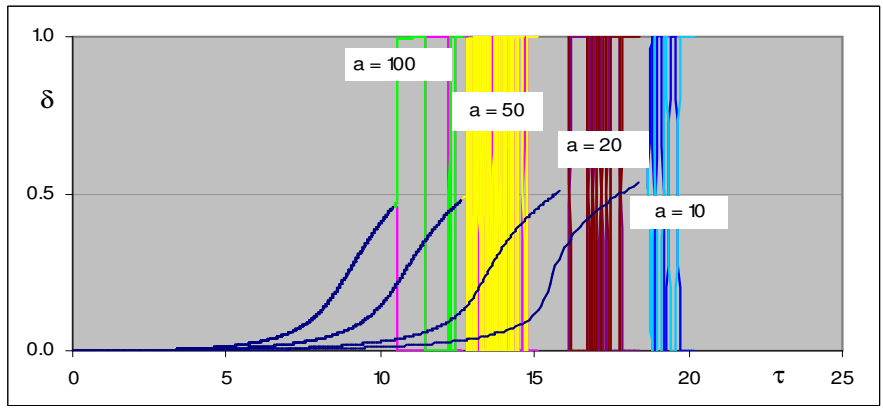

Fig.7. Shift from TdE at various $\alpha$, direct reaction A+B=AB, $\eta_d$=0.968, $\pi$=1.

COMPARISON TO CLASSICALLY SIMULATED AND EXPERIMENTAL RESULTS

We have encountered some qualitative similarities between the pictures in this paper and previous results of conventional thermodynamic simulation, obtained using *ad hoc* method of double compounds, detailed in [2].

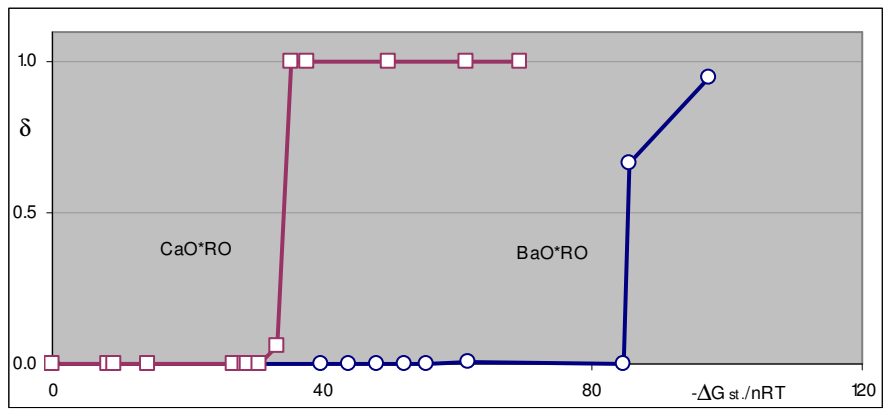

Fig.8. Classical thermodynamic simulation, ASTRA-4 [7], reaction (CaO(BaO)·RO+S), RO – various restricting oxides. To avoid divisions by zero, $\Delta G^0$ of MeO·RO formation from CaO/BaO and RO is plotted on abscissa instead of TdF.

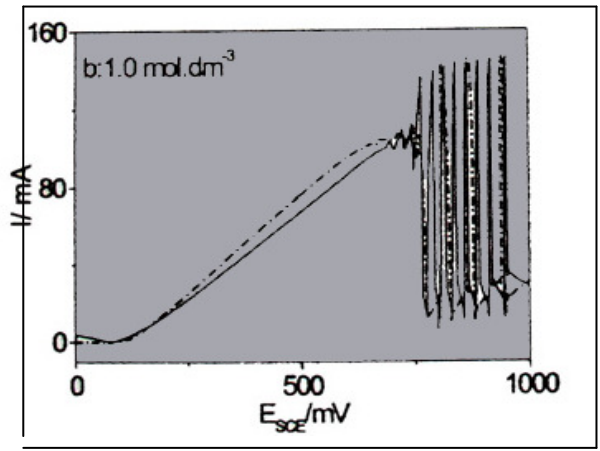

Fig.9. Electrical current oscillations, Cu electrode in trichloroacetic acid solution, concentration of CCl3COOH −1 M/dm³, adopted from [8].

Fig.8 features well pronounced TdE zones in reaction responses; actually, within those zones neither CaO nor BaO from the double oxides do not react with sulfur. As concerns to comparison between our results and experimental data, as far as we are aware of, nobody had set experiments in the way to draw results in τ-δ or TdF-δ reference frames. The growth factor τ is not a measurable value at all and can only be calculated (or preset during iterations). Many electrochemical publications contain the electrical current vs. voltage graphs, very similar to the graphical solutions to map (11), received in this work. Fig.9 shows the current oscillations of a copper electrode in trichloroacetic acid solutions in a cyclic experiment. The voltammetric curve with oscillations in that picture is strikingly similar to bifurcation diagrams in figures 3-5 and 7. It contains excellently expressed TdE, OpE and bifurcation areas, thus witnessing in behalf of our thermodynamic model. Indeed, electrical current runs through the electrochemical cell to restore equilibrium, from which the cell was shifted by external potential, and its value must be directly related, if not directly proportional to that shift. On the other hand, TdF, acting against electrochemical cell due to imposed voltage $E_j$, is linearly proportional to that voltage [5]. Many similar curves can be found, sometimes showing less perfect match with ours than the curves (direct and reverse) in Fig.9, perhaps due to not perfect experimental conditions.

CONCLUSION

This work presents a new approximation to basic map of discrete thermodynamics of chemical equilibria. It allows us to obtain more detailed pictures and more understanding of how the chemical system responds to external impact, uniting both previously found basic maps in one and covering much wider range of situations. Actual response bifurcation diagram is determined by a complex combination of basic system parameters – reaction stoichiometry, η, $ω_0$, and π.

Discrete thermodynamics of chemical equilibria has discovered the thermodynamically predicted oscillations in chemical systems. While bifurcations were found in many systems far enough from TdE in a wide set of the parameter values, the oscillations we observed occurred within relatively narrow limits. The shape of the simulated in this work bifurcation diagrams, including the shape of oscillations, has visibly changed and became closer to some experimental graphs, particularly in electrochemical systems. It doesn't mean, however, that bifurcation diagrams of the old type (like in Fig.1) are totally excluded.

The new map (11) ties together internal thermodynamic force and abstract external thermodynamic force, expressed via the Le Chatelier response in terms of the system deviation from isolated thermodynamic equilibrium. The map may be applied equally to closed and open systems.

"Chemical transformation" may be understood wider than just a chemical reaction: that could be any process of transforming something A into something B with changes in the system energy, e.g. the laser process of excitation A+$h\nu$=A* or its reverse, the same in photochemical reactions. If there exists a known relation between amounts of A and B (a kind of equilibrium ratio or constant), we don't have to know the transformation energy. That makes the DTd applications essentially more diversified than of the classical thermodynamics of chemical equilibria. Substantiation of the force in (9) will lead to individual maps with explicit dependencies of the system response on specific external impact parameters, like external potential, imposed onto electrical cell, and such maps will feature the same basic properties as general map (11).

Position of bifurcation point is very important: it is related directly to the external energy that must be supplied to the system in order to destabilize its thermodynamic branch. As it follows from the static diagrams in Fig.3 and the dynamic diagrams in Fig.6, left, the more robust is the system, i.e. the larger is $\eta$, the harder is the task to move it from TdE. The shift is the measure of the system response to external impact, while the thermodynamic equivalent of transformation $\eta$, also the system parameter, is the measure of the system robustness and resistance to external impact rather than $\Delta G^0$.